\documentclass[journal=jctcce,manuscript=article]{achemso}
\usepackage[breaklinks=true,colorlinks,citecolor=blue,linkcolor=blue,urlcolor=blue]{hyperref}
\usepackage{graphicx}
\usepackage{color}
\usepackage{amsmath}
\usepackage{amssymb}
\usepackage{siunitx}
\usepackage{bm}
\usepackage{subfig}
\usepackage{mhchem}
\usepackage[usenames,dvipsnames]{xcolor}

\setkeys{acs}{articletitle=true}

\usepackage[normalem]{ulem}

\newcommand{\HQS}{HQS Quantum Simulations GmbH, Rintheimer Strasse 23, 76131 Karlsruhe, Germany}

\title{Multi-level Protocol for Mechanistic Reaction Studies Using Semi-local Fitted Potential Energy Surfaces}

\author{Tomislav Piskor}
\email{tomislav.piskor@quantumsimulations.de, t.piskor@hotmail.de}
\affiliation{\HQS}
\alsoaffiliation{Theoretical Physics, Saarland University, 66123 Saarbr\"ucken, Germany}

\author{Peter Pinski}
\affiliation{\HQS}

\author{Thilo Mast}
\affiliation{\HQS}

\author{Vladimir V. Rybkin}
\affiliation{\HQS}

\begin{document}
	
	\begin{abstract}
		In this work, we propose a multi-level protocol for routine theoretical studies of chemical reaction mechanisms. The initial reaction paths of our investigated systems are sampled using the Nudged-Elastic Band (NEB) method driven by a cheap electronic structure method. Forces recalculated at the more accurate electronic structure theory for a set of points on the path are fitted with a machine-learning technique (in our case symmetric gradient domain machine learning or sGDML) to produce a semi-local reactive Potential Energy Surface (PES), embracing reactants, products and transition state (TS) regions. This approach has been successfully applied to a unimolecular (Bergman cyclization of enediyne) and a bimolecular (S$_\text{N}$2 substitution) reaction. In particular, we demonstrate that with only 50 to 150 energy-force evaluations with the accurate reference methods (here CASSCF and CCSD) it is possible to construct a semi-local PES giving qualitative agreement for stationary-point geometries, intrinsic reaction-coordinates and barriers. Furthermore, we find a qualitative agreement in vibrational frequencies and reaction rate coefficients. The key aspect of the method's performance is its multi-level nature, which not only saves computational effort but also allows extracting meaningful information along the reaction path, characterized by zero gradients in all but one direction. Agnostic to the nature of the TS and computationally economic, the protocol can be readily automated and routinely used for mechanistic reaction studies.
	\end{abstract}
	
	\maketitle
	
	\section{Introduction}
	Mechanistic studies of chemical reactions are one of the most important and wide-spread applications of computational quantum chemistry\cite{HRATCHIAN2005195}. A minimal meaningful workflow consists of locating reactants, products and a corresponding transition state (TS). This allows one to assess the thermodynamics and kinetics of reactions \emph{without} thermal contributions and thereby reveal reaction mechanisms. It is highly desirable to perform vibrational analysis for stationary points of a Potential Energy Surface (PES) to elucidate whether the structures correspond to minima or saddle points. This also allows to compute reaction rate coefficients via canonical Transition-State Theory (TST) taking vibrational degrees of freedom into account\cite{Truhlar_TST}. To be more rigorous, one should also compute reaction paths connecting the stationary points, one option being following the Intrinsic Reaction Coordinate (IRC) starting from the TS\cite{Fukui_IRC}.
	
	These objectives are routinely reached by direct dynamics approaches\cite{Truhlar_bimol}, \textit{i.e.} methods evaluating energy and its derivatives on-the-fly. The number of \textit{ab initio} quantum chemistry calls typically reaches hundreds and more. One should not ignore unsuccessful attempts to locate TS and find IRC, which are not uncommon in mechanistic reaction studies. Qualitatively accurate description of reaction paths involving breaking or forming of chemical bonds can only be achieved with advanced electronic structure methods. Often but not always those include static correlation, \textit{i.e.} multi-reference and multi-configurational wave function approaches\cite{pink_bible}. As an alternative, Density Functional Theory (DFT) with carefully selected hybrid and double-hydrid functionals can provide accurate results\cite{DFT_best_practices}. All these methods are, however, computationally demanding even for moderate-size systems. Therefore, the researchers often apply computationally cheaper methods for the mechanistic reaction studies, employing more accurate electronic structure theories only to stationary points (for an example, see the study of the reaction between ferrocenium and trimethylphosphine\cite{Chamkin_NEVPT2}). Some such heuristic method combinations are known as composite methods, or recipes, \textit{e.g.} the Gaussian-n family \cite{G4, G3XK, G4MP26X} and CBS-QB3 \cite{CBSQB3}. Despite many successful applications, these approaches imply properties inconsistent with geometries. They are designed not for PES exploration, but rather for stationary point calculations, and are prone to unpredictable errors (for an example, see \cite{CBSQB3_fail}). 
	
	Ideally, a composite method should provide accurate energies consistent with geometric structures at least in a relevant part of the configuration space, while being computationally feasible, \textit{i. e.} based only on few energy/force evaluations at the high-level of electronic structure theory. This implies obtaining a locally fitted PES: once it is available, one can complete many tasks ``free'' of charge, including reactive molecular dynamics simulations, canonical and variational TST calculations\cite{Truhlar_TST} and so on.
	
	A fitted PES can be efficiently generated using one of the rapidly evolving Machine-Learning (ML) approaches (for the general method overviews see review\cite{ml_ff} and perspective\cite{Behler}). These typically require little, if any, feature design: mainly, the structures with associated energies, and sometimes also energy gradients, are needed as input. They demonstrate remarkable flexibility, successfully describing even non-adiabatic processes\cite{Dral, Westermayr} and systems without a classical atomistic structural formula\cite{Lan2021}. The main type of application for such ML-PES is extensive Molecular Dynamics (MD) sampling (see exemplary applications to organic crystals\cite{Kapil_crystal} and liquid water\cite{Piaggi_water}). The resulting PES is often called ``global'' as it embraces vast regions of the configuration space, although the reactive regions of it are typically not covered. There are only a handful of applications of ML-PES fitting to reactive systems: second-order nucleophilic substitution (S$_\text{N}$2)\cite{Meuwly_SN2}, pericyclic\cite{Ang_peri}, decomposition\cite{Parrinello_decomposition}, dissociation\cite{Laage_dissociation}, Diels-Alder\cite{Duarte_DA} and proton-transfer reactions\cite{Meuwly_proton}. In addition, ML-PES have been successfully used for automatic mechanism discovery\cite{Friederich}. All these applications, however, aimed at ``global'' PES fitting required abundant data: typical sets include at least thousands of data points. Most of them, therefore, used Density Functional Theory (DFT) methods or cheaper many-body correlated wave functions as underlying electronic structure theory.
	
	To facilitate routine mechanistic reaction studies one should be able to construct semi-local reactive PES with only a few hundreds of data points using high-level \textit{ab initio} methods. Such work has been performed by Young \textit{et al.}\cite{Duarte_reactive} for several model processes, although the authors restricted themselves to DFT methods, which are typically insufficient to describe chemical reactions due to their single-reference character.
	
	In this work we propose a multi-level protocol for generating semi-local reactive PES for routine mechanistic reaction studies based on ML methods with small data sets and a combination of electronic structure theories. Computationally cheaper DFT is used to generate relevant structures, whereas high-level \textit{ab initio} methods are applied to refine the energetics. The multi-level nature of the protocol makes it suitable for incorporating evolving quantum computing methods which promise quantum advantage for correlated electronic-structure problems\cite{Aspuru_Guzik_ChemRev, Chan_ChemRev}, while remaining non-routine. The feasibility of the approach is demonstrated with two organic reactions: monomolecular Bergman cyclization of enediyne to para-benzyne\cite{Bergman_init} and a bimolecular S$_\text{N}$2 reaction of chloromethane with a bromide ion\cite{ingold1969structure}. The former is a textbook example of a chemical reaction involving multi-reference character\cite{Lindh_Bergman} and is treated with a complete-active-space self-consistent-field method, whereas the latter can be described with a single-reference correlated method\cite{Kekeres_SN2_CC} and is treated with a coupled-cluster approach.
	
	This paper is organized as follows. In section \ref{sct:methods}, we introduce the simulation protocol. In particular, we focus on obtaining relevant geometries for the investigated reactions, as well as reference methods and the machine learning technique of choice. Next, we present the results: PES sections, energy and force-prediction errors, and performance of the ML-PES for geometry optimization, vibrations and reaction rates. We conclude and give an outlook for this work in section \ref{sct:cao}.
	
	
	\section{Methodology}
	\label{sct:methods}
	
	\subsection{Simulation protocol}
	We applied the general protocol for fitting a semi-local reactive PES, which includes the following steps:
	\begin{enumerate}
		\item Optimize reactants and products with a cheaper electronic structure method;
		\item Find an approximate reaction path using the Nudged Elastic Band (NEB) method\cite{NEB} with a cheaper electronic structure method;
		\item Select points along the reaction path to form the data set;
		\item Calculate energies and forces with a reference correlated method;
		\item Split this data set into training, validation and test subsets;
		\item Fit the ML-PES, validate and test using the corresponding subsets.
	\end{enumerate}
	
	After being generated, the fitted PES was employed to compute the properties: stationary points, harmonic vibrations and intrinsic reaction coordinates in both directions from the TS. In addition, we calculated rate coefficients using transition state theory (TST). To evaluate the quality of the fitted semi-local PES the properties were compared with those obtained directly by the reference electronic structure method with energies and forces calculated on-the-fly.
	
	The details of each step are described below.
	
	\subsection{Electronic structure calculations}
	\label{sec:electronic_structure_calculations}
	We used DFT with the PBE exchange-correlation functional\cite{PBE} in combination with the double-zeta split-valence def2-SVP basis set\cite{def2basis} as a cheaper electronic structure method for both reactions. These calculations have been performed with \verb|NWChem| \cite{NWChem}.
	
	For the Bergman cyclization the complete active space self-consistent field (CASSCF)\cite{CASSCF_init} method with the double-zeta split-valence def2-SVP basis set\cite{def2basis} was used as the reference.
	For the initial configuration, we used MP2 natural orbitals to select the active space guess, which consisted of 12 electrons in 12 orbitals (12, 12) as suggested by Lindh and Persson \cite{Lindh_Persson}. The CASSCF wave function for atomic configurations in the data set was calculated with the previous guess for the molecular orbitals, so that no discontinuities occurred in the potential energy surface. CASSCF calculations were performed using \verb|PySCF|\cite{CASSCF_pyscf,sun2017}.
	
	The reference correlated method for the S$_\text{N}$2 reaction was coupled-cluster singles and doubles (CCSD)\cite{CCSD} based on the Hartree-Fock reference within the spin-restricted formalism. As in the previous example, CCSD calculations have been performed with \verb|PySCF| \cite{sun2017}. The basis set of choice has been the double-zeta correlation-consistent basis set cc-PVDZ\cite{cc_basis_1, cc_basis_3, cc_basis_9}. 
	
	\subsection{Data set generation}
	\verb|NWChem| \cite{NWChem} has been used to generate the data sets by performing NEB\cite{NEB} calculations. For both reactions, DFT (the cheaper method) energies and forces were used, with reactants as start points and products as end points. We used 100 beads for the S$_\text{N}$2 reaction and 200 for the Bergman cyclization. The geometries were randomly selected from the optimized elastic band. Thus, they lie on the approximate transition path of the PES obtained from the cheaper method, rather than on the path of the reference method.

	\subsection{Machine learning fit of the semi-local reactive PES}
	We opted for the symmetric gradient domain machine learning (sGDML) method\cite{chmiela2017,chmiela2018,chmiela2019} to fit the PES as it efficiently employs the forces and does not require feature design: sGDML uses inverse atomic distances as features and the interatomic forces as the output. Importantly, sGDML has been shown to faithfully reproduce PES from a limited number of structures. Moreover, it has been successfully applied to small molecules using coupled-cluster reference data\cite{sauceda2019}. 
	
	To perform one sGDML fit, a certain number of points from the data set is taken to define the training, validation and test set. These numbers are detailed in section \ref{sct:results}. The first step within one sGDML run is to generate a first model by constructing the kernel matrix out of the training points and setting values for the hyper-parameters $\sigma$ and $\lambda$. The regularization parameter is hereby fixed to a certain value and the length scale $\sigma$ is varied during the learning process. After the training step, the model is validated against the validation set, where the energy and gradient errors are determined. This procedure is continued until the best $\sigma$ is found. As a last step the optimized model is validated against the test set, which has no data from neither the training nor validation set and the errors in energy and gradients are compared.
	
	\subsection{Property calculation}
	We optimized the stationary points of the PES using DFT-optimized structures as initial guess and followed the IRC from the TS in both directions with \verb|pysisyphus|\cite{pysisyphus}. Both, the fitted PES and the reference electronic structure method in \verb|PySCF|\cite{sun2017} were employed as energy functions. Gradients for both PES types were calculated analytically.
	
	We calculated harmonic vibrations for all stationary points on the PES and computed the reaction rate coefficients using conventional TST with a harmonic oscillator-rigid rotor approximation for the partition functions. The working equations and the corresponding rates are given in the Supporting Information (Section 1).
	
	
	\section{Results}
	\label{sct:results}
	In this section, we present the results for our investigated examples. At first, the PES scans for the reference method and sGDML, as well as the energy differences between both methods will be shown. The next analysis will provide information about the mean absolute errors of the geometries between the reference and machine learning method for the reactant and TS. To perform geometry optimizations for both structures, we used the open-source package \verb|pysisyphus|. In addition, we computed and compared harmonic vibrations for the optimized structures. Besides the intrinsic reaction coordinate we finally compared the reaction rate coefficients for all methods and reactions. The intrinsic reaction coordinate has been determined with \verb|pysisyphus| as well, where the Euler-Predictor-Corrector integrator was used.
	
	\subsection{Bergman cyclization of enediyne}
	\label{subsct:enediyne}
\begin{figure}[H]
		\centering
		\includegraphics[width=1.00\textwidth]{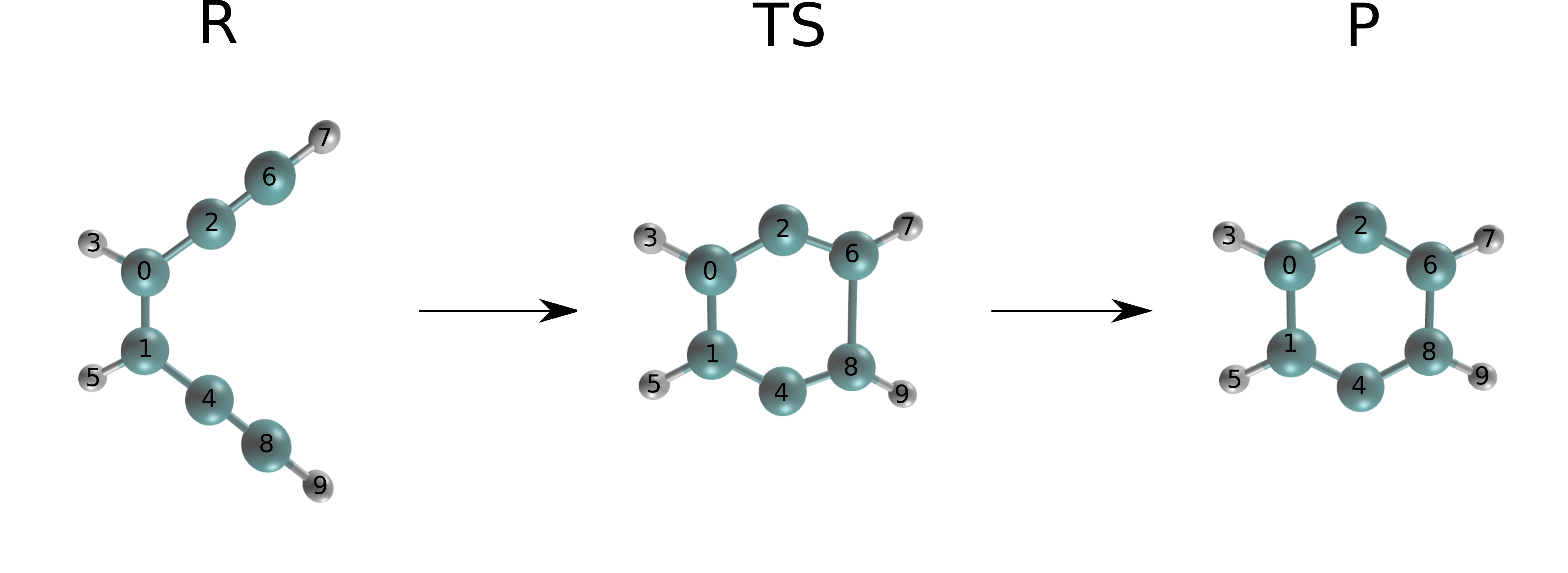}
		\caption{Bergman cyclization of enediyne. The reactant, transition state and product are labeled with R, TS, and P, respectively. Elements are colored as follows: carbon (cyan), hydrogen (white).}
		\label{fgr:Bergman}
	\end{figure}

	The reaction is shown in Fig. \ref{fgr:Bergman}, including the optimized reactant and product.
	
	\subsubsection{Model training}
	We created an sGDML model for the data set consisting of 200 data points: we used 150 training, 30 validation and 20 test points. The regularization parameter was set to $\lambda = 10^{-15}$ and the best length scale was found to be $\sigma = 6$. With these parameters, we obtained a model with a Mean Absolute Error (MAE) of \SI{0.0067}{kJ/mol} and a Root Mean Square Error (RMSE) of \SI{0.0071}{kJ/mol} for the energy, whereas the MAE and RMSE for the forces was \SI{0.0033}{kJ/{\AA} \cdot mol} and \SI{0.0075}{kJ/{\AA} \cdot mol}, respectively, on the test data set.
	
	\subsubsection{Potential energy surface scan}
\begin{figure}[h!]
		\centering
		\subfloat[Potential energy surface for CASSCF and sGDML. \label{fgr:enediyneallpes}]{\includegraphics[width=8cm]{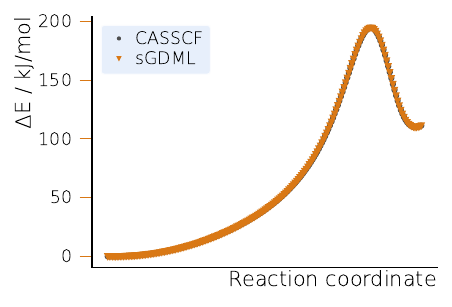}}\hfill
		\subfloat[Energy differences between the learned and CASSCF potential energy surface. \label{fgr:enediynediffpes}]{\includegraphics[width=8cm]{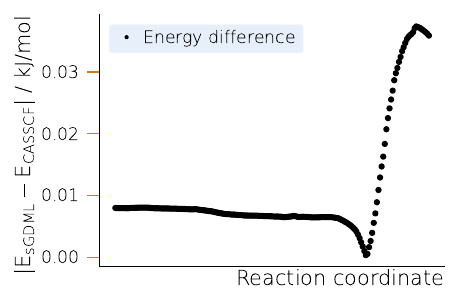}}
		
		\caption{Potential energy profile for the Bergman cyclization of enediyne along the optimized NEB path.}
		\label{fgr:enediynepes}
	\end{figure}

	The PES profiles corresponding to the DFT optimized NEB path computed with CASSCF and sGDML are shown in Fig.~\ref{fgr:enediyneallpes}. As expected from the small values of MAE and RMSE, the two surfaces agree within chemical accuracy (see Fig.~\ref{fgr:enediynediffpes}). Starting from the reactant state and moving towards the TS the energy error is small and practically constant. The smallest error can be found in the vicinity of the TS, increasing significantly towards the product state, although still being two orders of magnitude within chemical accuracy.
	
	\subsubsection{Geometry optimization}
\begin{figure}[htp]
		\centering
		\subfloat[Reactant state. \label{fgr:enediynereactant}]{\includegraphics[width=.29\textwidth]{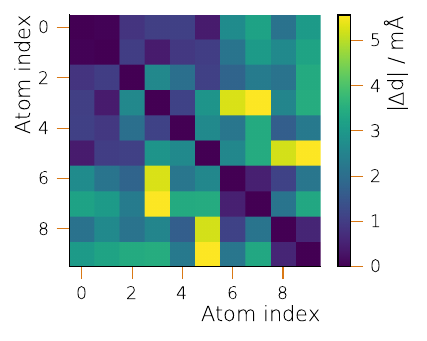}} 
		\subfloat[Transition state. \label{fgr:enediynets}]{\includegraphics[width=.31\textwidth]{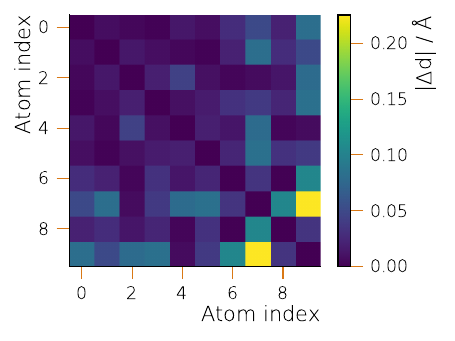}}
		\subfloat[Product state. \label{fgr:enediynep}]{\includegraphics[width=.31\textwidth]{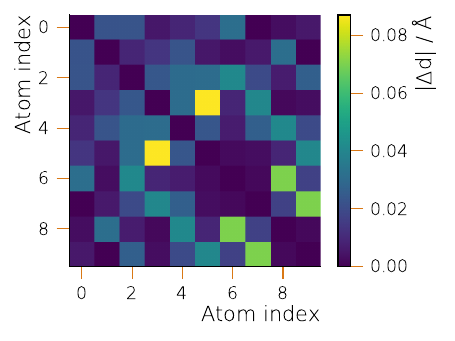}}
		
		\caption{Accuracy of geometric structures from the semi-local fitted PES for the Bergman cyclization of enediyne: position differences (MAE) for sGDML- and CASSCF-optimized structures. Atom numbering is given in Fig.~\ref{fgr:Bergman}.}
		\label{fgr:enediynegeometries}
	\end{figure}

	In Fig.~\ref{fgr:enediynegeometries} we compare the geometric structures of the stationary structures as obtained by sGDML and the reference method, CASSCF. Directly comparing the optimized geometries from both methods, we find that the MAE for all inter-atomic distances never reaches the value of $0.01$~{\AA} for the reactant state. However, for the TS, the error is considerably larger reaching a deviation of $0.22$~{\AA} in the worst case. Although it is a significant value, the corresponding distance is between two non-bonded hydrogen atoms separated by more than 5~{\AA}. For other distances, the error does not exceed $0.1$~{\AA}.
	
	\subsubsection{Vibrations, intrinsic reaction coordinates and reaction rate coefficients}
	
\begin{figure}[H]
		\centering
		\includegraphics[width=12cm]{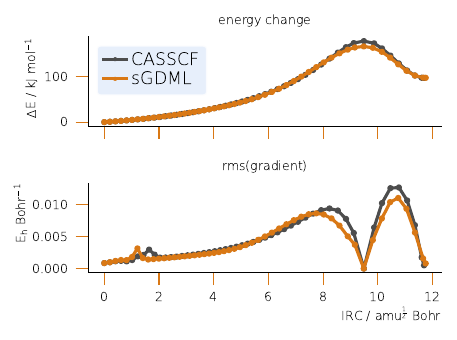}
		\caption{IRC for the Bergman cyclization of enediyne: relative energies and Root Mean Square gradients. The TS corresponds to the maximum energy value at approximately 9.5 value of the IRC displacement.}
		\label{fgr:enediyneall}
	\end{figure}

	After obtaining the optimized TS, we analyze the connection between it and the basins of reactants and products by integrating the IRC as shown in Fig.~\ref{fgr:enediyneall}. On the fitted semi-local PES, the TS does connect the reactants and products by a minimum energy path, which is in qualitative agreement with the reference CASSCF calculation along the entire curve as indicated by the maximum and RMS errors in the gradients.
	
	The successful computation of the IRC gives the first positive accuracy assessment of vibrational modes on the semi-local fitted PES: the imaginary frequency at the TS is needed to define the direction of the path. A more detailed look at vibrational frequencies reveals only semi-qualitative agreement between the sGDML and the reference CASSCF PES (see Table S1 in the SI): the differences between the frequencies reach up to approximately 200~\textrm{cm}$^{-1}$ for high-frequency nodes, which is still less than 10~\%.
	
	Rate coefficients dependent on structures, vibrations and reaction barriers are good integrated indicators of the fitted PES quality. The barrier heights for the Bergman cyclization are $\Delta E^\ddagger_\text{CASSCF} = \SI{194.59}{kJ}$ for the reference method and $\Delta E^\ddagger_\text{sGDML} = \SI{194.57}{kJ}$ for the fitted PES, which are in excellent qualitative agreement.
	Assuming $\textrm{T}=\SI{300}{K}$, we obtain the following reaction rate coefficients from the conventional TST (as described in the SI, Section 1) for the two methods: $k_\text{CASSCF} = \SI{3.4548e-22}{m^3 s^{-1}}$ for the CAS(12, 12) and $k_\text{sGDML} = \SI{4.3856e-22}{m^3 s^{-1}}$ for sGDML. The agreement is semi-qualitative and stems from the pre-exponential factors defined by the partition functions, which in turn depend on vibrations and structures being less accurate than the barriers.

	\subsection{S$_\text{N}$2 reaction of chloromethane with bromide}
	\label{subsct:sn2}
\begin{figure}[H]
		\centering
		\includegraphics[width=1.00\textwidth]{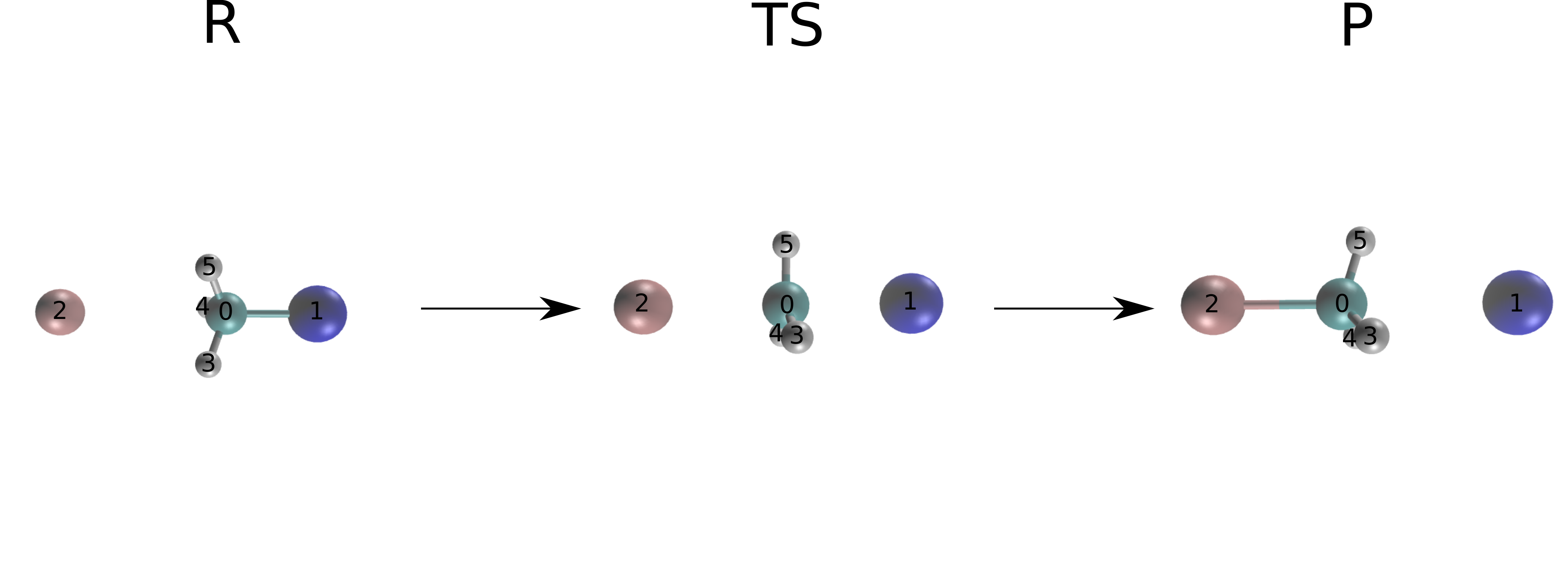}
		\caption{S$_\text{N}$2 reaction of chloromethane with bromide. The reactant, TS and product states are given as R, TS, and P, respectively. Elements are colored as follows: carbon (cyan), hydrogen (white), bromine (pink), chlorine (blue).}
		\label{fgr:sn2}
	\end{figure}

	The reaction between \ce{CH3Cl} and \ce{Br-} is shown in Fig. \ref{fgr:sn2}, including the optimized reactant and product.
	
	\subsubsection{Model training}
	To generate a machine learning model, we created a data set with a total of 100 geometries separated into 50 training, 30 validation and 20 test points. The regularization parameter was set to $\lambda = 10^{-15}$ and the best length scale was found to be $\sigma = 32$. With these parameters, we obtained a model with both the MAE and RMSE being \SI{0.0013}{kJ/mol}, whereas the MAE and RMSE for the forces were \SI{0.0105}{kJ/{\AA} \cdot mol} and \SI{0.0293}{kJ/{\AA} \cdot mol}, respectively, on the test data set.
	
	\subsubsection{Potential energy surface}
\begin{figure}[h!]
		\centering
		\subfloat[Potential energy surface for CCSD and sGDML. \label{fgr:sn2allpes}]{\includegraphics[width=8cm]{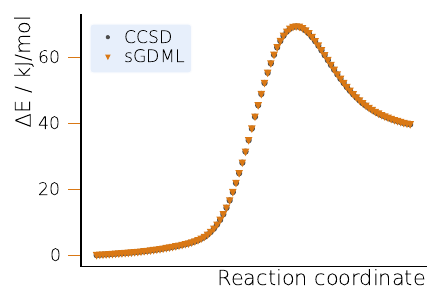}}\hfill
		\subfloat[Energy differences between the learned and CCSD potential energy surface. \label{fgr:sn2diffpes}]{\includegraphics[width=8cm]{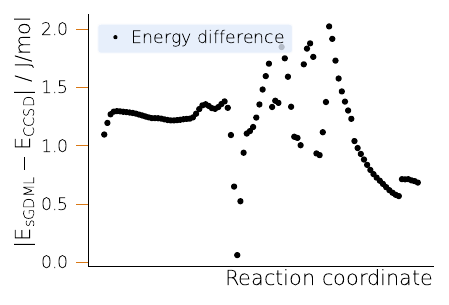}}
		
		\caption{Potential energy profile for the S$_\text{N}$2 reaction of chloromethane and bromide along the optimized NEB path.}
		\label{fgr:sn2pes}
	\end{figure}

	The potential energy profiles of the S$_\text{N}$2 reaction for the reference method, CCSD, and sGDML are shown in Fig.~\ref{fgr:sn2allpes}. The good agreement between CCSD and the sGDML fit along the (DFT) NEB path can be immediately seen in Fig.~\ref{fgr:sn2diffpes}. The energy difference is of similar order as for the Bergman cyclization and does not exceed 0.02~\textrm{kJ/mol}. However, we observe the higher errors in the TS region rather than in the vicinity of the product state.
	
	\subsubsection{Geometry optimization}
	\label{sct:sn2geomopt}
	
\begin{figure}[htp]
		\centering
		\subfloat[Reactant state. \label{fgr:sn2reactant}]{\includegraphics[width=.31\textwidth]{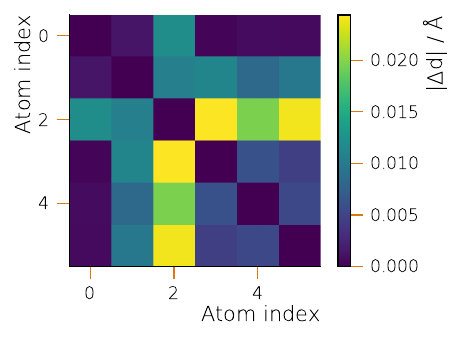}}
		\subfloat[Transition state. \label{fgr:sn2ts}]{\includegraphics[width=.29\textwidth]{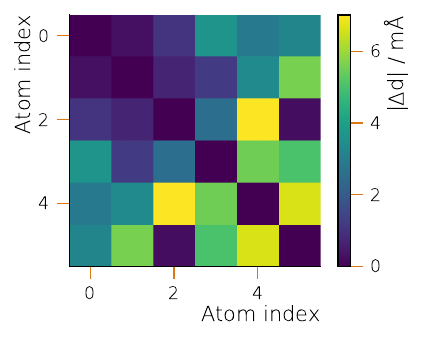}}
		\subfloat[Product state. \label{fgr:sn2p}]{\includegraphics[width=.31\textwidth]{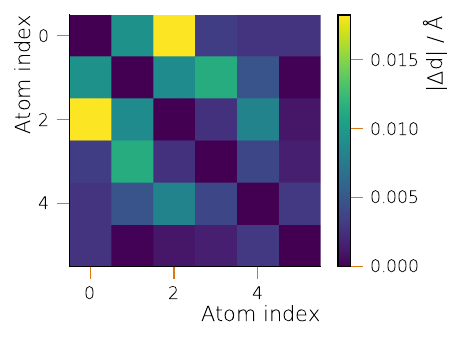}}
	
		\caption{Accuracy of geometric structures from the semi-local fitted PES for the S$_\text{N}$2 reaction of chloromethane and bromide: position differences (MAE) for sGDML- and CCSD-optimized structures. Atom numbering is given in Fig.~\ref{fgr:sn2}.}	
		\label{fgr:sn2geometries}
	\end{figure}

	In Fig.~\ref{fgr:sn2geometries} we compare the geometric structures of the stationary structures obtained by sGDML and the reference method. The general agreement between the reference and  fitted PES is better than for the Bergman cyclization, where the MAE never exceeds 0.025~{\AA} for a particular interatomic distance. For the S$_\text{N}$2 reaction, the better agreement is reached for the TS with the MAE staying below a value of 0.01~{\AA}. The largest deviation in case of the reactant state can be found for the distances between the non-bonded bromide ion (indicated by index 2) and the hydrogen atoms from the methylene group (indices 3, 4 and 5).

	\subsubsection{Vibrations, intrinsic reaction coordinates and reaction rate constants}
\begin{figure}[H]
		\centering
		\includegraphics[width=12cm]{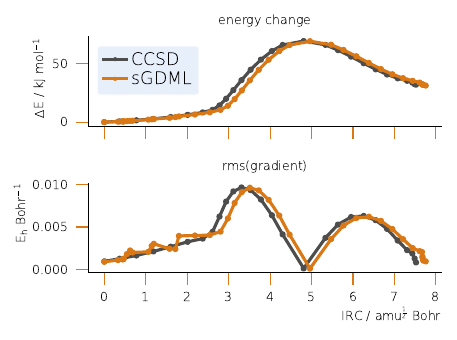}
		\caption{IRC for the S$_\text{N}$2 reaction of chloromethane with bromide: relative energies and and Root Mean Square gradients. The TS corresponds to the maximum energy value at approximately 4.8 value of the IRC displacement.}
		\label{fgr:sn2all}
	\end{figure}

	After obtaining the optimized TS, we analyze the connection between it and the basins of the reactants and products by integrating the IRC as shown in Fig.~\ref{fgr:sn2all}. On the fitted semi-local PES, the TS does connect the reactants and products by a minimum energy path, which is in qualitative agreement with the reference CCSD calculation along the entire path as indicated by the maximum and RMSE in the gradients.
	
	As in the previous example, successful computation of the IRC indicates the correct Hessian structure at the TS with the only imaginary eigenvalue corresponding to the same eigenvector as in the reference method. A more detailed look at vibrational frequencies reveals only qualitative agreement between the sGDML and the reference CCSD PES (see Table 2 in the SI): the differences between the frequencies reach several hundreds of \textrm{cm}$^{-1}$ for both high- and low-frequency modes.
	
	The barrier heights for the S$_\text{N}$2 reaction were identical up to the second digit: $\Delta E^\ddagger = \SI{69.47}{kJ}$ for CCSD and  the fitted PES. At $\textrm{T}=\SI{300}{K}$, the CCSD reaction rate coefficient is $\SI{1.8187e-31}{m^3 s^{-1}}$, whereas this value is $\SI{2.2287e-32}{m^3 s^{-1}}$ for sGDML, which is an order of magnitude smaller than the reference method. This difference occurs due to the larger deviations in the harmonic vibrations, defining the preexponential factors in the TST equations.
	
	\section{Discussion}
	\label{sct:disc}
	The most important property of the fitted semi-local PES surfaces obtained by the proposed protocol is the general stability with respect to PES exploration techniques. Indeed, for both unimolecular and bimolecular reactions geometry optimizations have converged to the stationary points, connected by physically meaningful minimum energy paths. This is achieved by the multi-level nature of our approach.
	
	On the one hand, the NEB driven by the cheaper method (here DFT with a PBE exchange-correlation functional) generates structures relatively close to the reactive region of the system as defined by more accurate reference electronic structure methods (here, CASSCF or CCSD). Indeed, DFT methods are known to predict sensitive geometric structures\cite{Grimme_DFT} (even if the energy barriers are not precise). This is illustrated by the S$_\text{N}$2 reaction studied in this work: in Fig. S1 of the SI we see that the highest energy point on the NEB path for both PBE and CCSD is located at the similar values of the reaction coordinate, although the energy barriers differ dramatically.
	
	On the other hand, structures sampled by NEB using the cheaper method should be far enough from the minimum energy path of the reference method. The points on the true minimum energy path (approximated by NEB\cite{NEB}) have only one non-zero gradient component -- the one along the path tangent direction\cite{Fukui_IRC}. Keeping in mind that sGDML uses gradients as inputs for fitting, providing only structures on the minimum energy path would provide no information about the nature of the PES along orthogonal directions and make model very sensitive to numerical noise. Consequently, using a cheaper electronic structure method for the NEB simulation is not only computational effort saving, but also essential for the quality of the data set used to train a gradient-based ML model.
	
	Despite qualitative agreement achieved by the semi-local fitted PES in structures, energy barriers and IRC, a more subtle property, vibrational spectrum, is computed less accurately. Although the structure of Hessian is qualitatively correct, some frequencies can differ by several hundreds of \textrm{cm}$^{-1}$. This is particularly noticeable for the S$_\text{N}$2 example and as a consequence leads to a significant error (order of magnitude) in the reaction rate coefficient as compared to the reference value. This effect must have to do with a smaller number of training points (only 50) used for the model as compared to the Bergman cyclization (150 points). Another reason is the fact that the relevant PES region for the S$_\text{N}$2 is flatter as this for the Bergman cyclization (compare Fig.~\ref{fgr:enediyneall} and Fig.~\ref{fgr:sn2all}), one indication of which is the lower energy barrier for the former reaction. This makes finite-difference evaluation of the second derivatives of energy less reliable and requires a better PES sampling for obtaining qualitative results.

	\section{Conclusions and Outlook}
	\label{sct:cao}
	In this work, we have proposed a multi-level protocol for reaction mechanism studies aiming to match the accurate electronic structure theory description at the reduced computational cost. The approach involves cheaper electronic structure method to generate structures along the reaction path via the NEB method, evaluating energies and gradients for a restricted set of them with an accurate (reference) theory and fitting a semi-local reactive PES using machine learning method, sGDML. The fitted PES is then used for computing properties.
	
	The protocol has been applied to a unimolecular (Bergman cyclization) and a bimolecular (S$_\text{N}$2) reactions using PBE/CASSCF and PBE/CCSD, respectively, as cheap/reference method pairs, the results being compared with those obtained on-the-fly by the reference method. Our approach achieves quantitative agreement in structure optimization for the PES stationary points, reaction energy barriers and IRC following using as little as 50-150 training points (corresponding to energy and gradient calculations with the reference method), whereas the agreement in vibrational frequencies and reaction coefficients is only (semi-)qualitative. The key to the performance of the protocol is its multi-level character as the differences between the reference and cheap methods are essential to provide meaningful information about the reactive PES region.
	
	The results are encouraging as important objective of mechanistic reaction study can be performed with high-level electronic structure method without prior knowledge of the TS structure (due to the application of the NEB) requiring only scarce amount of data. The protocol is simple and can be automatized for routine calculations.
	
	At the same time, we have not considered more complex reactions with multiple steps, product branching and shallow minima. These cases would require more subtle approaches to sampling and larger amounts of training data.
	
	Our simple protocol can be further improved by applying smart selection of training points using molecular fingerprints\cite{fingerprints} and using methods even computationally cheaper than DFT (such as modern tight-binding, GFN2-XTB\cite{XTB}) for sampling structures. Moderately accurate reference methods applied here should be substituted by those providing qualitatively correct PES, such as multi-reference approaches, to match the experimental accuracy. Furthermore, the potential of different ML fitting techniques should be explored.

	\section{Conflicts of interest}
	The authors declare no conflicts of interest.
	
	\section*{Acknowledgements}
	This work was supported by the German Federal Ministry of Economic Affairs and Climate Action through the PlanQK project (01MK20005H) and the AQUAS project (01MQ22003A).
	
	\newpage

	\bibliography{ml-paper}
	
\end{document}